\documentclass[11pt,a4]{article}
\usepackage{graphicx,fleqn,times,caption2}
\begin{document}
\renewcommand{\captionfont}{\small}
\large
\vskip .1in
\centerline{\large\bf On Bars, Haloes, their Interaction and 
their Orbital Structure}
\vskip .5in
\renewcommand{\thefootnote}{\alph{footnote}}
\centerline{E. ATHANASSOULA}
\vskip .1in
\centerline{\it Observatoire de Marseille Provence}
\vskip .02in
\centerline{\it 2, Place Le Verrier}
\vskip .02in
\centerline{\it 13248 Marseille C\'edex 04}
\vskip .02in
\centerline{\it France}
\vskip .02in
\centerline{\it email: lia@oamp.fr}
\par\noindent
\vskip 1.in
\noindent
{\bf Abstract : }
A live halo plays an active role in the formation and evolution of
bars by participating in the angular momentum redistribution which
drives the dynamical evolution. Angular
momentum is emitted mainly by near-resonant material in the bar region
and is absorbed mainly by near-resonant 
material in the halo and in the outer disc. This exchange determines
the strength of the bar, the decrease of its pattern speed, as well as
its morphology. Thus, contrary to previous beliefs, a halo can 
help the bar grow, so that bars growing in galaxies with responsive
massive haloes can become stronger than bars growing in disc dominated
galaxies. During the evolution the halo does not stay axisymmetric. It
forms a bar which is shorter and 
fatter than the disc bar and stays so all through the simulation, 
although its length grows considerably with time. I discuss the
orbital structure in the disc and the halo and compare it with
periodic orbits in analytical barred galaxy potentials. 
A central mass concentration (e.g. a central black hole, or a central
disc) weakens a bar and increases its pattern speed. The effect of the
central mass concentration depends strongly on the model, being  
less strong in models with a massive concentrated halo and a strong bar.
\vskip .2in


\section{Introduction}

Bars are elongated structures seen in the central parts of a large
fraction of disc galaxies. Their morphological, photometrical and
kinematical properties have been widely studied. Unfortunately, there
is no review of observational results which is both complete and up
to date, but the reader can consult Kormendy (1982); Bosma (1992);
Buta, Crocker \& Elmegreen (1996); Athanassoula, Bosma \& Mujica
(2002) and Block et al. (2005) for
earlier reviews or reviews covering sub-topics.

Bars are very common features. Eskridge et al. (2000), using 
a statistically well-defined sample of 186 disc galaxies from the Ohio
State University Bright Spiral Galaxy Survey, find that 56\% are
strongly barred in the H band, while another 16\% are weakly
barred. Grosb{\o}l, Patsis \& Pompei (2002), using a smaller sample of
53 spirals observed in 
the K band, find that about 75\% of them have bars or ovals. 

In this paper I will discuss a number of results I have obtained
recently on the formation and the dynamical evolution of bars. In
particular, I will discuss the effect of angular momentum exchange
within the galaxy, the role of the halo, 
the orbital structure in barred galaxies and the effect of a
central mass concentration on the evolution.

\section{\bf Orbital structure in barred galaxies}

The first step towards understanding the dynamics of a given structure is
to understand its main families of periodic orbits. Indeed, if a
periodic orbit is stable, it can trap around it regular orbits of
similar orientation and morphology. On the
other hand, unstable periodic orbits introduce chaos. A growing body
of evidence shows that chaotic orbits can
indeed be a considerable fraction of the total and that they contribute
significantly to the morphology and to the kinematics of bars
(e.g. Binney 1982; Athanassoula et al. 1983; Pfenniger 1984b; Teuben \&
Sanders 1985; Hasan \& Norman 1990; Patsis, Athanassoula \& Quillen 1997; 
Contopoulos 2002; El-Zant \& Shlosman 2002, 2003). 

\subsection{\bf Periodic orbits}

\begin{figure} 
\begin{center}
\rotatebox{0}{\includegraphics[width=35pc]{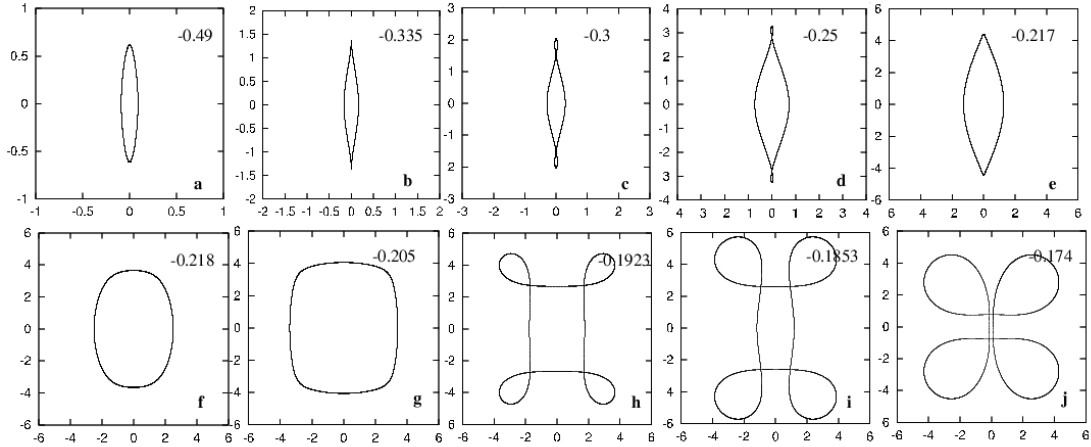}}
\end{center}
\caption[]{ Examples of x$_1$ orbits. The bar is oriented along the
  $y$ axis and has a semi-major axis equal to 6 length units. The
  Jacobi constant, given in the upper right corner of 
  each panel, increases from left
  to right and from top to bottom. From Skokos, Patsis \&
  Athanassoula 2002, Monthly Notices of the Royal Astronomical,
  Society, 333, 847, Blackwell Publ.}
\label{fig:x1}
\end{figure}

The basic families of periodic orbits were calculated first in 2D models. 
Due to restricted computing power, the earlier studies were rather
limited (e.g. de Vaucouleurs \& Freeman 1972, Michalodimitrakis
1975). The first extensive study was done by Contopoulos \&
Papayannopoulos (1980),
who used a very simplified potential to calculate
orbits both inside and outside corotation (hereafter CR). They called
the main family x$_1$ and showed that within CR it is elongated along the
bar. They also presented the banana-like Lagrangian orbits, around the
Lagrangian points $L_4$ and $L_5$, the family x$_2$ of central orbits
elongated perpendicular to the bar and the retrograde family x$_4$. 
Athanassoula et al. (1983) extended this work using a much more
realistic bar potential, namely the Ferrers' potential (Ferrers
1877). Unfortunately their nomenclature is different from that of
Contopoulos \&
Papayannopoulos (1980), the main family being called here
$B$, the perpendicular one $A$ and the retrograde one $R$. They
confirmed that the $B$ (x$_1$) family is the backbone of 
the bar. They used surfaces of section to show that most regular
quasi-periodic orbits are trapped around family $B$ (x$_1$) or around the
main retrograde family $R$ (x$_4$). They also showed that more massive
and/or more eccentric bars introduce more chaos.

Figure~\ref{fig:x1} (from Skokos, Patsis \& Athanassoula 2002a) shows
a sequence of orbits of the x$_1$ family. Following them in order of
increasing 
Jacobi constant, we see a morphological sequence, first discussed by
Athanassoula (1992). Namely the orbits first become cuspy at the
apocenter, where, for yet larger energies, they acquire two loops, one
at each apocenter. At yet higher energies these two loops disappear
and the orbits become oval-like and 
then rectangular-like. At the largest energies the orbits form four
loops, one at each of the four corners of the rectangular-like shape. 

More than half of the orbits displayed in figure~\ref{fig:x1} close
after one revolution around the center and two radial
oscillations. They are thus resonant 1:2 orbits and are often referred
to a such. At higher energies, however, the orbits close after one
rotation and 4 (or more) radial oscillations. Such orbits are often
considered as members of the x$_1$ family, but can also be called 1:4,
1:6, or in general 1:$n$, orbits.  

\begin{figure}
\begin{center}
\rotatebox{0}{\includegraphics[width=30pc]{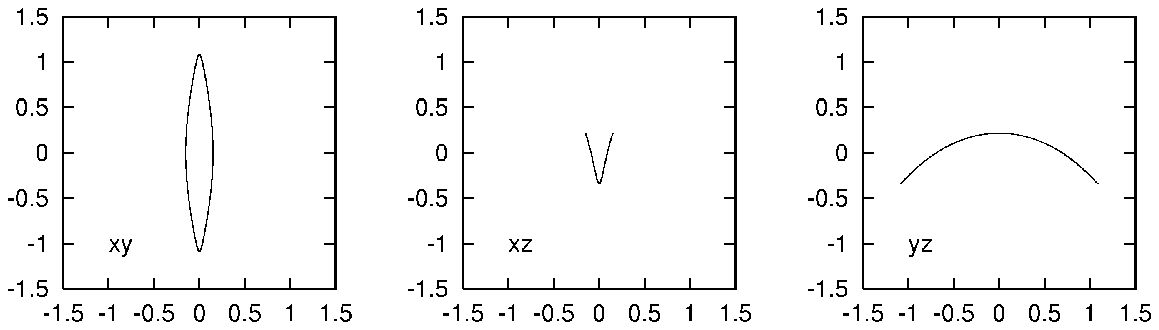}}
\vskip 10pt
\rotatebox{0}{\includegraphics[width=29pc]{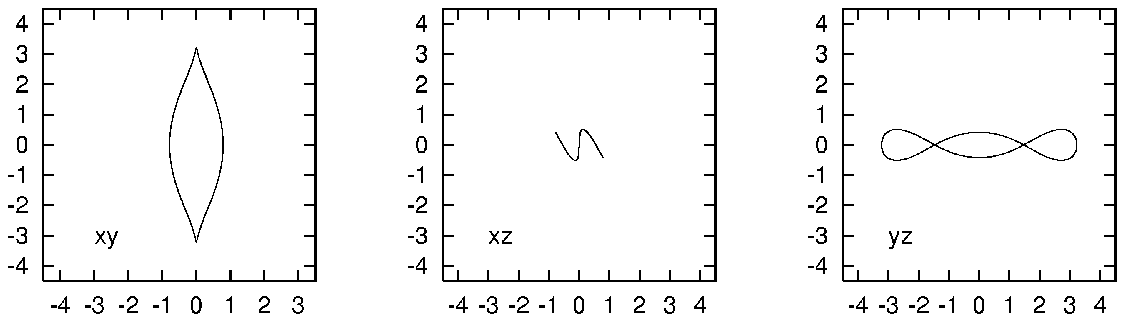}}
\vskip 10pt
\rotatebox{0}{\includegraphics[width=29pc]{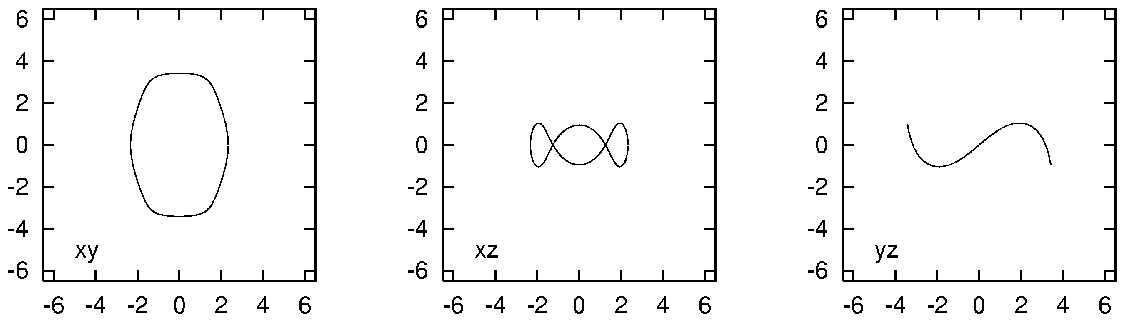}}
\vskip 10pt
\rotatebox{0}{\includegraphics[width=29pc]{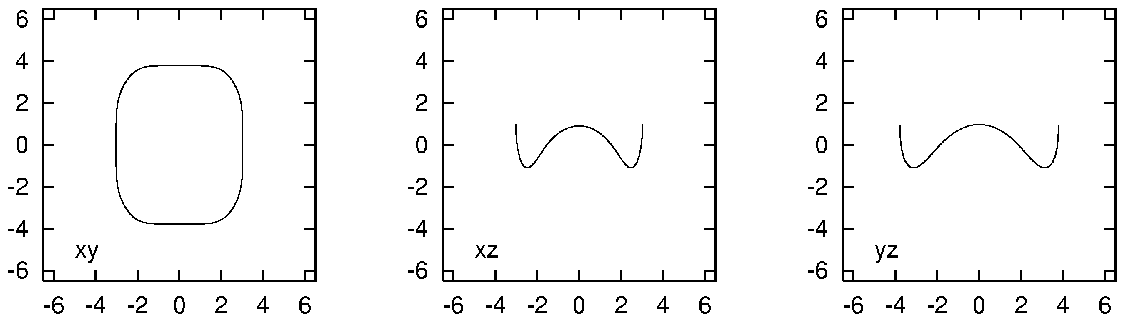}}
\vskip 10pt
\rotatebox{0}{\includegraphics[width=30pc]{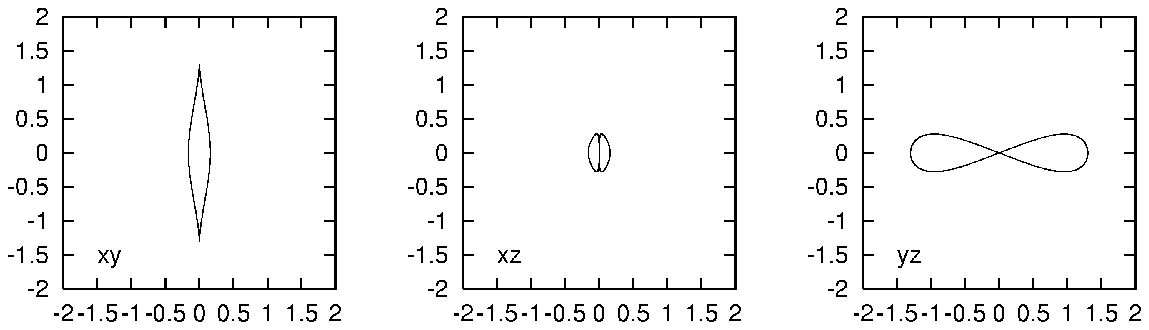}}
\end{center}
\caption[]{ Three orthogonal views of characteristic 3D orbits in
  barred galaxies. The bar is oriented along the $y$ axis and has a
  semi-major axis equal to 6 length units. From top to 
  bottom, these orbits are members of the 
  families x1v1, x1v3, x1v4, x1v5 and x1v2. }
\label{fig:3Dorbits}
\end{figure}

3D studies showed that the orbital structure is much richer and more
complicated. Pfenniger (1984a) initiated such studies and
showed that there are
several families of 3D orbits, bifurcated at the vertical resonances
of the main planar family. This work was supplemented and extended
in a series of four papers (Skokos, Patsis \& Athanassoula 2002a, b;
Patsis, Skokos \& Athanassoula 2002, 2003). 
I will briefly recall here some results from these papers that
are relevant to the subjects discussed here. I will
also follow their notation.

The backbone of 3D bars is the x$_1$ tree, i.e. the x$_1$
family plus a tree of 2D and 3D families bifurcating from it 
(Skokos, Patsis \& Athanassoula 2002a, b). These 3D families are called
x1v$n$, where $n$ is the order with which they are bifurcated in the
fiducial model. Thus the family which bifurcates at the lowest energy
is x1v1, followed by x1v2, then x1v3 etc. Characteristic orbits from
these families in the fiducial model of Skokos, Patsis \&
Athanassoula (2002a) are given in 
figure~\ref{fig:3Dorbits}. The first four (from families x1v1, x1v3,
x1v4 and x1v5) are stable. The last one (from family x1v2) is unstable.
A comparison of figures \ref{fig:x1} and \ref{fig:3Dorbits}
clearly shows the morphological similarity of the orbits of the x$_1$
family in the 2D problem and the ($x$,$y$) projection of the x1v$n$
orbits. On the other hand, the x1v$n$ orbits extend well out of the
equatorial plane. It is thus the trapping around the periodic orbits
of the x$_1$ tree that will define the shape of the bar in the
equatorial plane, as well as its thickness perpendicular to it. 

Skokos, Patsis \& Athanassoula (2002a) found also 3D families of
banana-like orbits around the 
$L_4$ and $L_5$ Lagrangian points. Considerable sections of the
families are stable. More surprising, they also found {\it stable} periodic
orbits around the unstable Lagrangian points $L_1$ and $L_2$. The
family is planar, starts unstable, but turns stable at larger energy
values.

\subsection{Chaos and how to measure it}

Galaxies are systems that contain both order and chaos, i.e. they have
both ordered and chaotic orbits. For 2D systems for which there is a
rotating frame of reference in which the energy is an integral of the
motion, 
the most straightforward way of distinguishing between the two is to
use surfaces of section. Even idealised galaxies, 
however, often do not fulfill these conditions. Thus many methods for
measuring chaos have been so far proposed (see Contopoulos 2002 for a
review). 

\begin{figure}
\begin{center}
\includegraphics{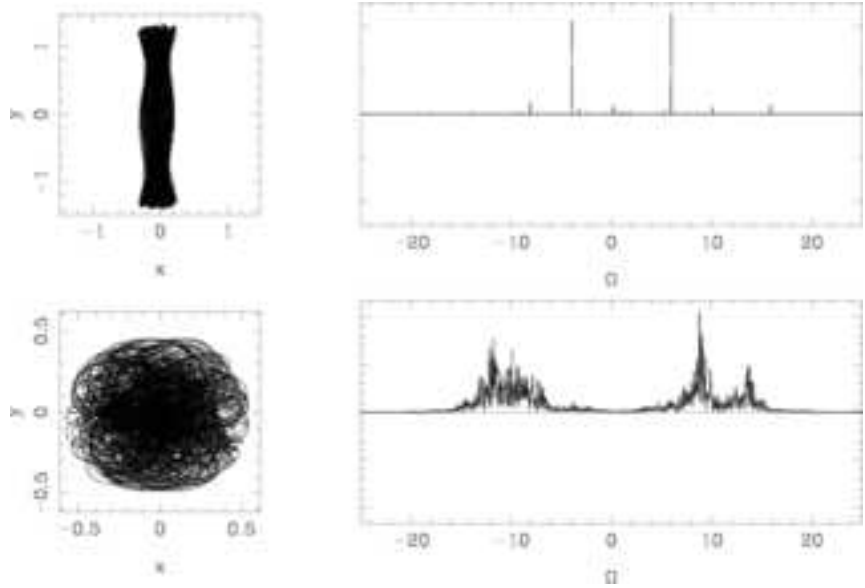} 
\end{center}
\vspace{7.1cm}
\caption{The ($x$,$y$) projection of two orbits (left panels) and the
  amplitude of the corresponding power spectra of $x + i y$ (right
  panels). The orbit described in the upper panels is regular, while the one in
  the lower panels is chaotic.
}
\label{fig:spectra}
\end{figure}

I will here use a method proposed by the person we are honouring in
this meeting, Henry Kandrup, and his collaborators (Kandrup, Eckstein
\& Bradley 1997). By
Fourier transforming a quantity related to the orbital coordinates,
e.g. the cylindrical radius $R$, or the complex quantity $x + i y$,
one obtains a spectrum. If
the orbit is regular, all the power of the spectrum lies in few,
well defined peaks, while, if the orbit is chaotic, 
there is a continuum of frequencies within which the power is
distributed. This is illustrated in
figure~\ref{fig:spectra}, which shows two orbits, one regular and the
other chaotic, and their corresponding spectra. 

\begin{figure}
\begin{center}
\includegraphics{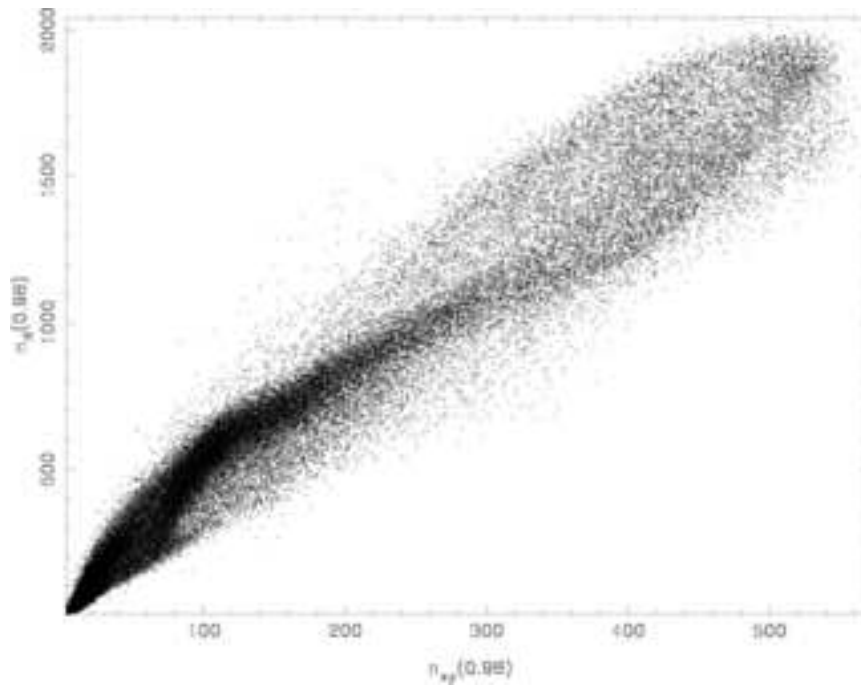} 
\end{center}
\vspace{8.5cm}
\caption{Complexity of disc orbits obtained from the $R$ spectrum
  as a function of that obtained, for the same orbits, from the $x+iy$
  spectrum.  
}
\label{fig:chaosRxy}
\end{figure}

Kandrup et al. (1997) define as the complexity $n(k)$ of an orbit, or,
more precisely, of an orbital segment,  the number of frequencies in its
discrete Fourier spectrum that contain a fraction $k$ of its total
power. Regular orbits will have small values of $n(k)$, as opposed to
chaotic orbits which will have big values. The optimum value for $k$
depends on the size of the orbital 
segment and the frequency of the sampling. For the cases I discuss in
this paper, I use potentials from $N$-body simulations and
consider 65536 time steps in 40 bar rotations. For such cases,  
Misiriotis \& Athanassoula (in
preparation) found $k$ = 0.98 to be the optimum value. Other large values
would have also been reasonable choices, since the Rank-Order
correlation coefficient 
of the complexities obtained using $k$ values between 0.90 and 0.99 is
always larger than 0.9 (Misiriotis \& Athanassoula, in
preparation). This confirms the result found by Kandrup et al. (1997)
for a different potential and under different conditions. Misiriotis \&
Athanassoula (in preparation) also found a strong correlation between
the complexity found from the $x + i y$ spectrum and that calculated
from the $R$ spectrum. This is illustrated in figure~\ref{fig:chaosRxy},
for the disc orbits of a simulation with a strong bar.
The Rank-Order correlation coefficient of these two quantities
is, in this case, 0.99. Kandrup et al. (1997) found strong
correlations of the complexity with the  
short term Lyapunov exponents, often used to measure chaos. They thus
concluded that the complexity $n(k)$ is a robust quantitative
diagnostic of chaos. It is particularly straightforward to
apply to $N$-body simulations and it executes very fast, so that it
can be applied to a very large number of orbits and simulations. I
will thus adopt it for the estimates presented here.  
This definition shares the shortcoming of other chaos
definitions. i.e. there is no clear dividing line between chaotic
orbits and regular, but very complex ones. 

\section{The effect of the halo} 
\label{sec:halo}

$N$-body simulations of the early seventies (Miller, Prendergast \&
Quirk 1970; Hohl 1971) already showed that bars form
spontaneously in galactic discs. At that time the observational
evidence for the existence of dark haloes around individual galaxies
was hardly compelling, so the discs in these simulations are
self-gravitating. Only a few years later haloes were propelled
into the center of scientific discussions. Ostriker \& Peebles (1973) 
were the first to check the effect of a heavy
halo on the bar instability and found it to be stabilising. Although
the number of particles in their simulations did not exceed 500, their
work is very insightful. They introduced the parameter $t_{op}$, which
is the ratio of kinetic energy of rotation to total gravitational
energy and they concluded that halo-to-disc mass ratios of 1 to 2.5 and an
initial value of $t_{op} = 0.14 \pm 0.03$ are required for
stability. Several later papers (e.g. Efstathiou, Lake \& Negroponte
1982; Athanassoula \& Sellwood 1986; Bottema 2003)
confirmed the stabilising tendency of the halo. Yet, as we will see in the
next section, this is an artifact, due to the fact that these
simulations were either 2D, or had a rigid halo, or had too few
particles. Thus the halo was not properly described and stabilised the
bar. The first doubts about an entirely passive role of the halo
component were voiced by Toomre (1977).

The importance of a live halo in order to model correctly the
evolution of a barred galaxy was clearly demonstrated by Athanassoula
(2002, hereafter A02). Two simulations are compared in this paper.
They have initially identical disc components and their haloes have 
initially identical mass distributions. However, in one of the two
simulations the halo is live, i.e. it is composed of particles, while
in the other it is rigid, i.e. it is an imposed potential. Thus in the
former simulation the halo can absorb angular momentum, while in
the latter it can not. The difference in the evolution is very
striking. The simulation with the live halo grows a very strong bar,
which, when seen side-on, has a strong peanut shape. On the contrary, the
simulation with the rigid halo has a very mild oval in the innermost
regions, and hardly evolves when seen edge-on. The large difference
between the results of the two simulations 
argues strongly that the angular momentum absorbed by the halo
can be a decisive factor in the evolution of the bar component.

\begin{figure}
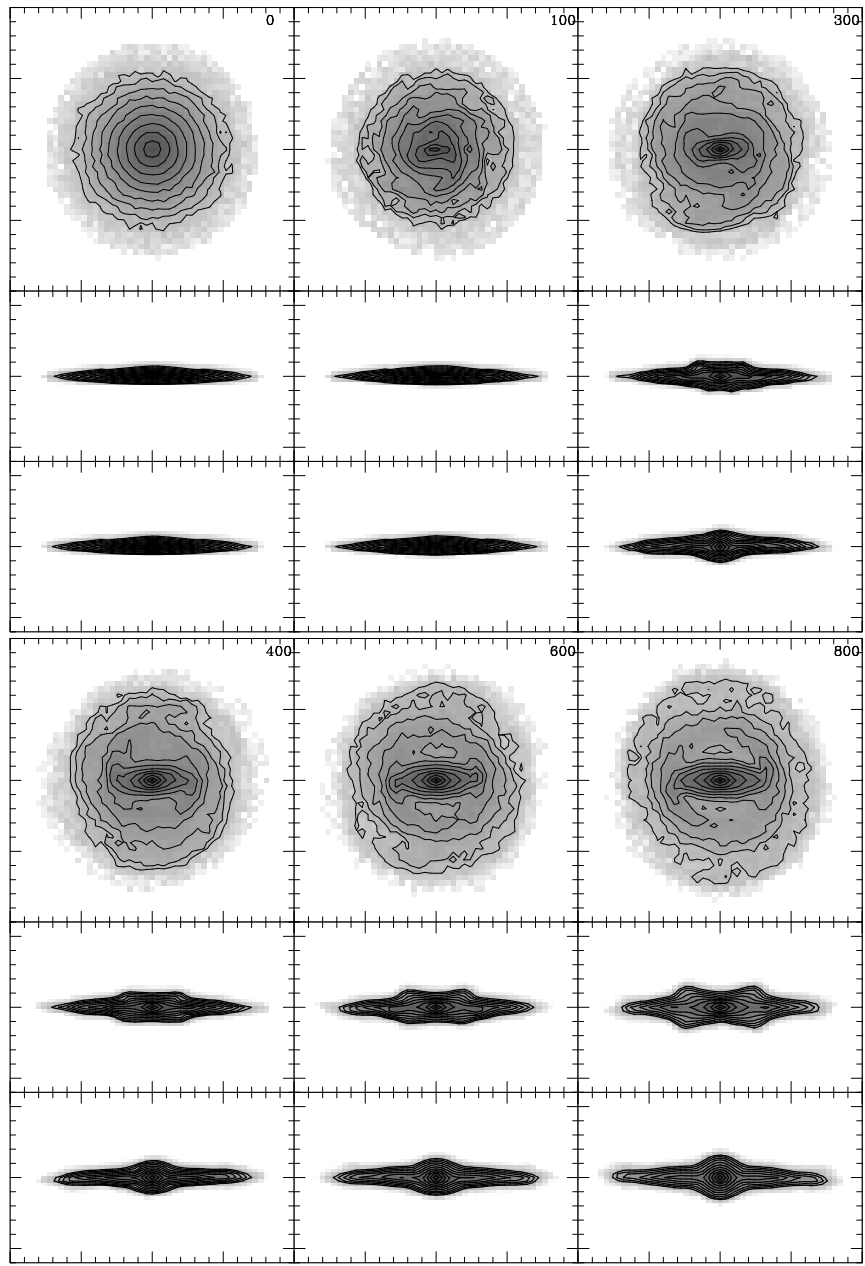

\begin{center} 
\rotatebox{0}{\includegraphics[width=27pc]{evol1_MH.ps}}
\rotatebox{0}{\includegraphics[width=27pc]{evol2_MH.ps}}
\caption[]{Formation of a bar in an initially axisymmetric disc. The
  model is of MH-type. The upper and fourth rows give 
  the face-on views; the second and fifth ones the side-on views and
  the third and sixth rows the end-on views. Time
  increases from left to right and from top to bottom and is given in
  the upper right corner of each face-on panel. Here and elsewhere in 
  this paper, times are given in computer units as defined by 
  Athanassoula \& Misiriotis (2002).}
\label{fig:evol-MH}
\vskip -2.0cm
\end{center}
\end{figure}

\begin{figure} 
\begin{center} 
\centering\rotatebox{0}{\includegraphics[width=27pc]{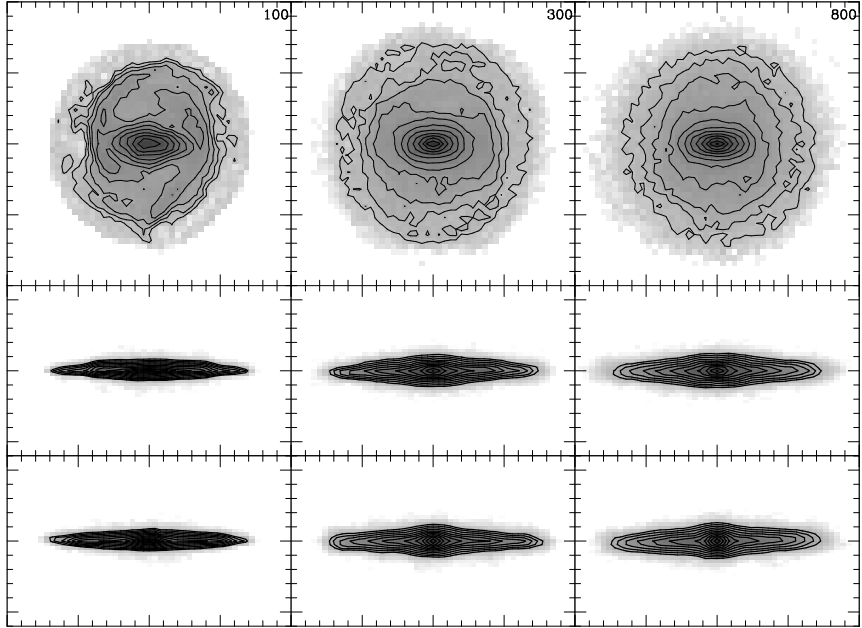}}
\caption[]{Formation of a bar in a disc dominated model (MD-type). The
  layout is as in Fig.~\ref{fig:evol-MH}. }
\label{fig:evol-MD}
\end{center}
\end{figure}

Athanassoula \& Misiriotis (2002, hereafter AM02) studied the
morphology, photometry and kinematics of the 
bar as a function of the central concentration of the halo. They found
that the bars that grow in centrally concentrated haloes are stronger,
longer and thinner than bars in less centrally concentrated
haloes, as can also be seen in figures~\ref{fig:evol-MH} and
\ref{fig:evol-MD}. AM02 called the former MH-types and the latter MD-types. 
MH-type bars grow initially slower than MD-types, in good
agreement with what was found by previous studies (e.g. Athanassoula
\& Sellwood 1986). They reach, however, higher amplitudes. Their
face-on 
isodensity contours are rectangular-like, while the corresponding ones
in MD-types are more elliptical-like. Their $m$ = 4, 6 and even 8 Fourier
components of the density are well out of the noise and their
amplitude reaches a considerable
fraction of the corresponding $m$ = 2, contrary to MD-types in which the
$m$ = 6 and 8 are negligible. The density profile along the bar
major axis (face-on view) also differs in the two types 
of bars. In MH-types it is rather flat, with an abrupt drop at the end
of the bar, while in MD-types it drops near-exponentially with
distance from the center (figure 5 in AM02). Bars in
MH-type models often have ansae and/or an inner ring, which is
elongated, but not far from circular and has the same major axis as
the bar, as inner rings observed in barred galaxies (Buta 1986). Their 
side-on\footnote{The side-on view is the the edge-on view in which the
  line of sight is along the bar minor axis. The end-on view is the
  edge-on view in which the line of sight is along the bar major axis.} 
shape evolves first to boxy and then to peanut or X shape, in contrast
to the MD-types for which the side-on outline stays boxy. The side-on
velocity field of MH-types shows cylindrical rotation, while the
MD-types do not. More information on these properties can be found in
AM02. 

It is worth noting (see also Athanassoula 2003b)
that a number of the properties of the MH-type galaxies are found in
early type bars. Thus, early type bars are longer than late type ones
(Elmegreen \& Elmegreen 1985). They often have ansae (Sandage 1961) and
flat projected light profiles 
along the bar major axis, in contrast to late type bars
which have more sharply falling profiles (Elmegreen \& Elmegreen
1985; Ohta, Hamabe \& Wakamatsu 1990; Ohta 1996). Strong early-type
bars have rectangular-like outlines (Athanassoula et al. 1990)
and $m$ = 6 and 8 Fourier components of the density of
considerable amplitude, contrary to late type bars, where these
components are negligible (Ohta 1996).  

From the results summarised in this section, it becomes clear that,
contrary to previous beliefs, a {\it live} halo can 
{\it help the bar grow}, so that bars growing in galaxies with responsive
massive haloes can become stronger than bars growing in
disc-dominated galaxies. This marks the end of a paradigm.

\section{Angular momentum exchange}
\label{sec:theory}

The secular growth of the bar component in isolated galaxies is driven
by the angular momentum redistribution within them.  
This was initially proposed, for galaxies with no spheroidal
component, by Lynden-Bell \& Kalnajs (1972), and later extended to
galaxies with haloes and/or bulges by Athanassoula (2003, hereafter A03).
Angular momentum is emitted mainly by near-resonant
stars in the inner disc (namely at inner Lindblad resonance,
hereafter ILR), or at higher order inner resonances (1:$n$) and, if the
perturbation is growing, also by non-resonant stars in that region. It
is absorbed mainly by stars at near-resonance in the outer disc -- 
CR and outer Lindblad resonance (hereafter OLR) -- and at all
resonances of the halo. The latter effect of the halo can be 
followed analytically if the distribution function of this component
depends on the energy only. A 
full analytical treatment of other distribution functions is not yet
available, but a preliminary analysis of models with non-isotropic
velocity distribution
(Fuchs \& Athanassoula, in preparation) indicates that these
could drive an even stronger bar growth.  

The net result is a transport of the angular momentum outwards. 
Colder material can emit/absorb more angular
momentum that hotter material. Thus the halo is less responsive than the disc
per equal amount of resonant mass. There is, however, not much material
in the outer disc, where the density is very low, while the halo can
be very massive. It can thus be that the halo absorbs more angular
momentum than the outer disc, and this has proven to be the case in
many $N$-body simulations, as will be discussed in the next
section. Since the bar is a negative angular momentum
`perturbation' (Kalnajs 1971; Lynden-Bell \& Kalnajs 1972), by losing
angular momentum it becomes stronger. 

The Lynden-Bell \& Kalnajs (1972) formalism and its extension to
include spheroidal components (A03) were able to predict successfully
which resonances emit and which absorb angular 
momentum. Extending it, however, beyond such qualitative results to obtain
a quantitative estimate of the bar growth, is not possible, 
since, as stressed by Weinberg (2004), 
it assumes that the perturbations grow slowly over a very long time and 
that transients can be ignored. For this reason, quantitative
estimates of the bar growth have, so far, only been obtained numerically. 

Tremaine \& Weinberg (1984) also studied the effect of the resonances,
focusing on the effect of the angular momentum exchange on
the bar pattern speed. They
showed that the dynamical friction on the bar arises from 
stars which are near-resonant with the rotating bar. They derived an
analogue of Chandrasekhar's formula (Chandrasekhar 1943) for spherical
systems, valid when the angular velocity of the bar does not change too
slowly. Weinberg (1985) calculated that the angular momentum exchange
between the bar and the halo will
cause a considerable slow-down of the former within a few bar
rotations. As we will see in the next section, bars in $N$-body
simulations also 
present such a slow-down (Little \& Carlberg 1991a, 1991b; Hernquist
\& Weinberg 1992; Athanassoula 1996; Debattista \& Sellwood 2000;
A03;O'Neill \& Dubinski 2003; Valenzuela \& Klypin 2003), in good
qualitative agreement with the analytical results. A quantitative
comparison may not be meaningful, because of the limitations
underlying the theory, in particular the fact that theory has not
treated so far both the change of pattern speed and the bar growth
simultaneously.   

\begin{figure}[h] 
\begin{center}
\rotatebox{-90}{\includegraphics[width=20pc]{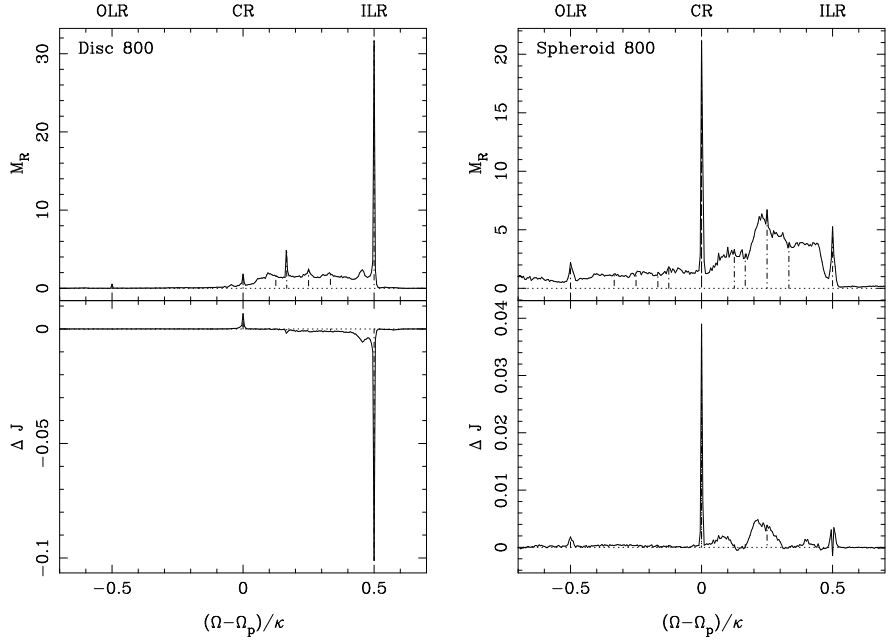}}
\end{center}
\caption[]{Resonances in the disc and the spheroidal component. The
  upper panels give, for the time $t$ = 800, the mass per unit
  frequency ratio, $M_R$, as a function of that ratio. The frequency
  ratio is defined as $(\Omega - \Omega_p) / \kappa $, the ratio of the
  angular frequency in the frame in which the bar is at rest to the
  epicyclic frequency. The lower
  panels give $\Delta J$, the angular momentum gained or lost by
  particles of a given frequency ratio between times 500 and 800, as
  a function of that frequency ratio, calculated at $t$ = 800. The
  left panels correspond to the disc component and the right ones to
  to the spheroid. The vertical dot-dashed lines give the positions
  of the main resonances. 
}
\label{fig:resonances}
\end{figure}

\section{Results from $N$-body simulations}
\label{sec:simul}

Contrary to real galaxies, $N$-body simulations are well suited for
studying the angular momentum exchange within a galaxy. This has been
one of the main goals of A02 and A03
and I will retrace here a number of the steps made in those papers.

A02 and A03 first checked that there is a considerable amount of
near-resonant material in the halo component. This can also be seen,
for a strong bar simulation, in the upper panels of
figure~\ref{fig:resonances}, where I plot the mass per unit frequency
ratio, $M_R$, as a function of the frequency ratio $(\Omega -
\Omega_p) / \kappa$. Here $\Omega$ is the angular frequency of the
orbit, $\kappa$ is its radial frequency and $\Omega_p$ is the bar pattern
speed. It is clear that the distribution is not uniform, and that its peaks
are located at the main resonances. In
all simulations, the disc has a strong peak at ILR, which is made of
particles trapped around this resonance and constituting the backbone
of the bar. Secondary peaks can be found
at other resonances -- as e.g. inner 1:3, inner 1:4, CR or OLR -- whose
existence and height varies from one simulation to another. More
important, the halo component also shows similar peaks. The highest is
at CR, while secondary peaks can be seen at ILR and OLR. Such peaks
can be seen in all simulations I analysed, again with varying
heights. 

The bottom panels of figure~\ref{fig:resonances} show the way the
angular momentum is exchanged. For the disc component it is emitted 
from the region within the bar, and particularly the ILR, and absorbed
at CR (and in some simulations also at OLR). However, the amount of
angular momentum emitted is much more than what the outer disc
absorbs. This is understood with the help of the bottom right panel,
which shows that all the halo resonances absorb a considerable amount
of angular momentum, much more so than the outer disc. A more complete
discussion and analysis can be found in A02 and A03. Thus simulations
confirm the angular momentum exchange mechanism suggested by the
analytical work and show that the halo can be an important agent
in this respect. 

This also explains why strong bars can not grow in simulations with
rigid haloes. Indeed, as initially discussed in A02, in such cases the
halo can not take angular momentum from the bar and thus it limits bar
growth. The difference between the bar strength in live and in rigid haloes
should be larger in cases when the role of the halo in the angular
momentum exchange is more important. 

The global redistribution of angular momentum was followed in
Debattista \& Sellwood (2000), A03, O'Neill \& Dubinski (2003) and
Valenzuela \& Klypin (2003) and shows clearly that angular momentum is
taken from the disc by the halo. This is in good agreement both with
the theoretical predictions and with the above more detailed analysis.

Simulations also show that the bar grows as a result of the angular
momentum exchange, as expected from theory. As already discussed in
the previous section, they show that the bar grows particularly strong
if the halo can take from it considerable amounts of angular
momentum. For such simulations, A03 found a clear 
correlation between the amplitude of the bar and the angular momentum
taken by the halo.

Simulations also show that the bar slows down as a result of the angular
momentum exchange, as expected from theory (Little \& Carlberg 1991a,
1991b; Hernquist 
\& Weinberg 1992; Athanassoula 1996; Debattista \& Sellwood 2000;
A03; O'Neill \& Dubinski 2003; Valenzuela \& Klypin 2003).
Thus one would expect an anti-correlation between the strength and the
pattern speed of the bar, and this was indeed established by A03, for
a large number of simulations.  

Theoretical arguments predict that the angular momentum emitted or
absorbed at a given resonance depends not only on the density of
matter there, but also on how cold the near-resonant material is. 
This is borne out by the simulations (A03). Indeed, if the disc is hot
(i.e. has a high initial $Q$) and/or the halo is very hot, then the
bar does not grow to be very strong and does not slow down much, even
in cases where the halo is very dense.
Examples of this are given by A03.

The angular momentum exchange also determines the morphology of the
bar (Athanassoula 2005). Simulations where only 
little angular momentum has been exchanged
harbour either an oval or a very short bar. Ovals are mainly found in
simulations with hot discs, while short bars are predominantly found
in simulations with hot haloes. At the other extreme, in simulations
in which a large amount of angular momentum is redistributed the bar is
strong, resembling the MH model of AM02 (see also
figure~\ref{fig:evol-MH}). As already discussed in  
section ~\ref{sec:halo}, such bars resemble the strong bars in early type
barred galaxies. Thus one can argue that a considerable amount of
angular momentum has been redistributed in such galaxies between the
disc and the spheroidal component. This would be partly taken by the
strong bulge component these galaxies have, and partly by their halo.
Extreme cases of such galaxies are some examples of bar dominated 
early type discs presented by Gadotti \& de Souza (2003), in which the 
disc is not a major component any more, since 
a very considerable fraction of its mass is now within the bar component.

I have so far taken into account only the disc and halo (and sometimes
bulge) 
components. Yet the complete picture of angular momentum
exchange can be more complicated. Galaxies (particularly late types)
have also a gaseous disc component. This may give angular momentum to
the bar, and thus decrease its strength (Berentzen et al. 2004). Furthermore,
galaxies are not isolated universes, and thus can interact with their
companions, or satellites. If the latter absorb angular momentum, then the
bar can grow stronger than in the isolated disc (Berentzen et al.
2004). This is 
in good agreement with observational results that show that more bars
can be found in interacting than in isolated galaxies (Elmegreen,
Elmegreen \& Bellin 1990). 

In isolated galaxies, angular momentum can only be redistributed
between the various components, i.e. there should be as much angular
momentum absorbed as emitted. Thus a very massive and responsive halo that
can absorb large amounts of angular momentum is not sufficient to
ensure important angular momentum redistribution. This can indeed be
limited by the amount of angular momentum that the inner disc can
emit, so that, if the disc has a very low surface density and/or is
very hot, little angular momentum will be exchanged. In other words, it
is not useful to increase the capacity of the absorbers if the
emitters do not follow, and vice versa. Similar arguments can constrain
the position of CR. If this is located in the inner parts of the disc
it will privilege the absorbers to the detriment of the emitters. This
occurs in simulations where the halo absorption is limited so that the
contribution of the outer disc is essential. On the contrary, a CR
located in the outermost parts of the disc favours emitters. This
ensures a maximum exchange in cases where the halo is capable of
absorbing a lot of angular momentum.  
 
\section{A bar in the halo component}

\begin{figure}
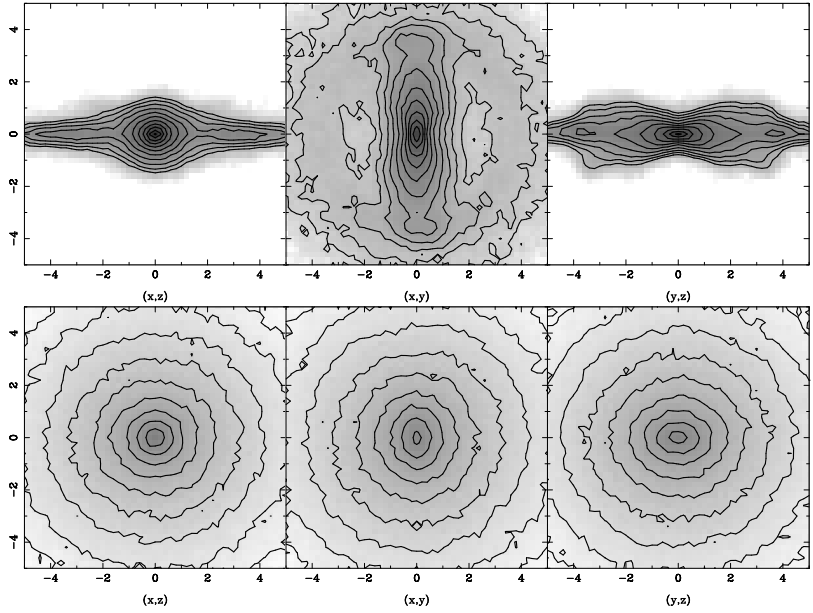
 
\begin{center}
\rotatebox{-90}{\includegraphics[width=9.5pc]{3views_disc.ps}}
\rotatebox{-90}{\includegraphics[width=9.5pc]{3views_halo.ps}}
\end{center}
\caption[]{ Three orthogonal views of the disc (upper panels) and halo
components (lower panels). The central panel is a face-on view, while
the two others are edge-on; side-on for the right panels and end-on for
the left ones. Note that the halo component does not stay axisymmetric, 
but forms an oval in its inner parts, which I call the
halo bar. }
\label{fig:halo-disc}
\end{figure}

The halo evolves dynamically together with
the disc. In simulations in which a considerable amount of angular
momentum has been exchanged and which have thus formed a strong bar,
the halo does not 
stay axisymmetric. It also forms a bar, or more precisely an oval,
which I will call, for brevity, the halo bar to distinguish it from
the bar in the disc component, which I will refer to as the disc
bar. An example is seen in 
figure~\ref{fig:halo-disc}, which compares the morphology of the disc
bar (upper panels) to that of the halo bar (lower panels). A very
clear case of such a structure is also seen in 
figure 2 of Holley-Bockelmann, Weinberg \& Katz (2003).

\begin{figure} 
\begin{center}
\rotatebox{-90}{\includegraphics[width=18pc]{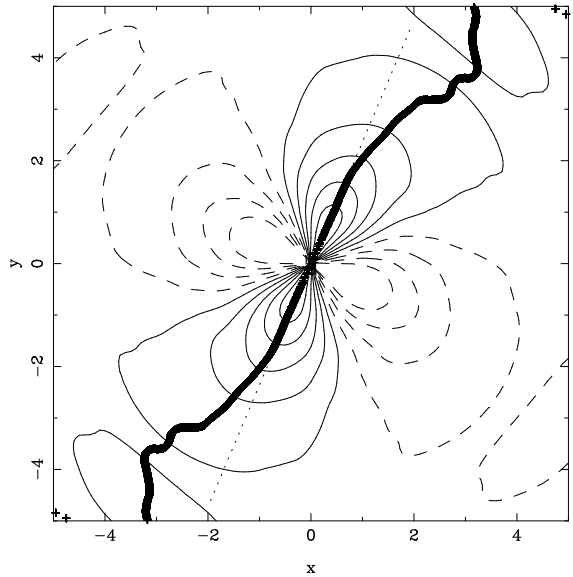}}
\end{center}
\caption[]{ Isocontours of the $l$=2 $m$=2 component of the halo mass
  distribution on the equatorial plane. Positive isocontours are given
  with solid lines and negative ones with dashed lines. The thick
  line shows the phase of the halo bar and the thin dotted line gives
  the position angle of the disc bar.}
\label{fig:isocl2m2}
\end{figure}

Halo bars are
triaxial, but nearer to prolate than to oblate, with their minor axis
perpendicular to the disc equatorial plane. Their axial ratio in this plane
(ratio of minor to major axis) is considerably larger than that of the
corresponding 
disc bar. It increases with increasing radius, so that halo bars tend
to become axisymmetric in the outer parts. Since the change in axial
ratio 
is very gradual, it is not easy to define precisely the end of the
halo bar, and thus to calculate its length, so that any measurement
will have a considerable error. It is clear, however, that it is
always considerably shorter than the disc bar. 

\begin{figure} 
\begin{center}
\rotatebox{-90}{\includegraphics[width=20pc]{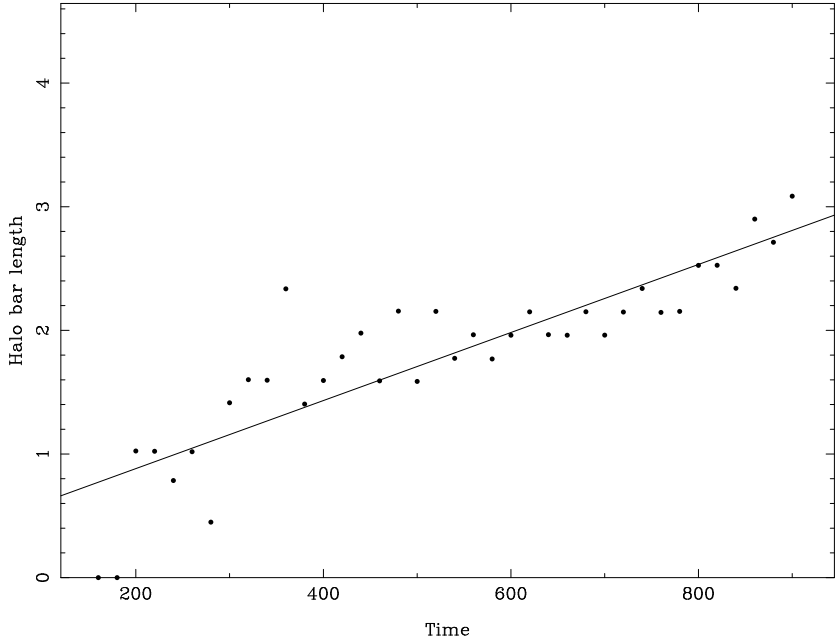}}
\end{center}
\caption[]{ Length of the halo bar as a function of time. The solid
  line is a least square fit. }
\label{fig:hbar-time}
\end{figure}

The phase and amplitude of the halo bar can be best studied with the
help of a decomposition into spherical harmonics. The mass
distribution of the $l$ = 2 and $m$ = 2 component on the equatorial
plane, as obtained from such a decomposition (Athanassoula, in
preparation), is given in figure~\ref{fig:isocl2m2}. It shows clearly
that in the inner parts, where the phase of the halo bar does not
change much with radius, the halo bar has roughly the same orientation
as the disc bar. This is true at all times after both bars have grown
sufficiently to allow a relatively accurate measurement of their
amplitude and phase. Closer examination, however, shows that the halo
bar lags 
the disc bar slightly, by something like a couple of degrees in the
innermost parts. This shift is always trailing and increases with
increasing distance from the center.  Thus the $m$=2 component of the
halo continues well outside the halo bar, trailing behind the disc
bar, so that, after a certain
distance, the structure can be described as a trailing very open spiral. 

The length of the halo bar can be defined by setting a limiting value
for the phase difference between the halo and the disc bar. It is
given as a function of time for a simulation with a strong disc bar
in figure~\ref{fig:hbar-time}. The
spread in the measurements reflects the difficulty of defining
precisely the end of the halo bar. Yet a least square fit shows
clearly that the length of the halo bar increases with
time. Comparison with similar data, but now for the disc bar, shows
that the length of the halo bar increases slower than
that of the disc bar. 

I carried out such an analysis for a number of simulations in which
the halo bar is sufficiently strong so that its properties can be
measured accurately. I find that the halo bar length
correlates with the disc bar length and that the regression line
changes with time, as would be expected since the disc bar grows 
faster than the halo bar. I found also correlations between
the disc and the halo bar strengths, as well as between the
ellipticity of the halo bar and its flattening towards the equatorial
plane. More information on the halo bar properties and their relation with the
disc bar properties will be given elsewhere (Athanassoula, in preparation).

\section{Orbital structure in the disc and halo components}

\begin{figure} 
\begin{center}
\rotatebox{0}{\includegraphics[width=20pc]{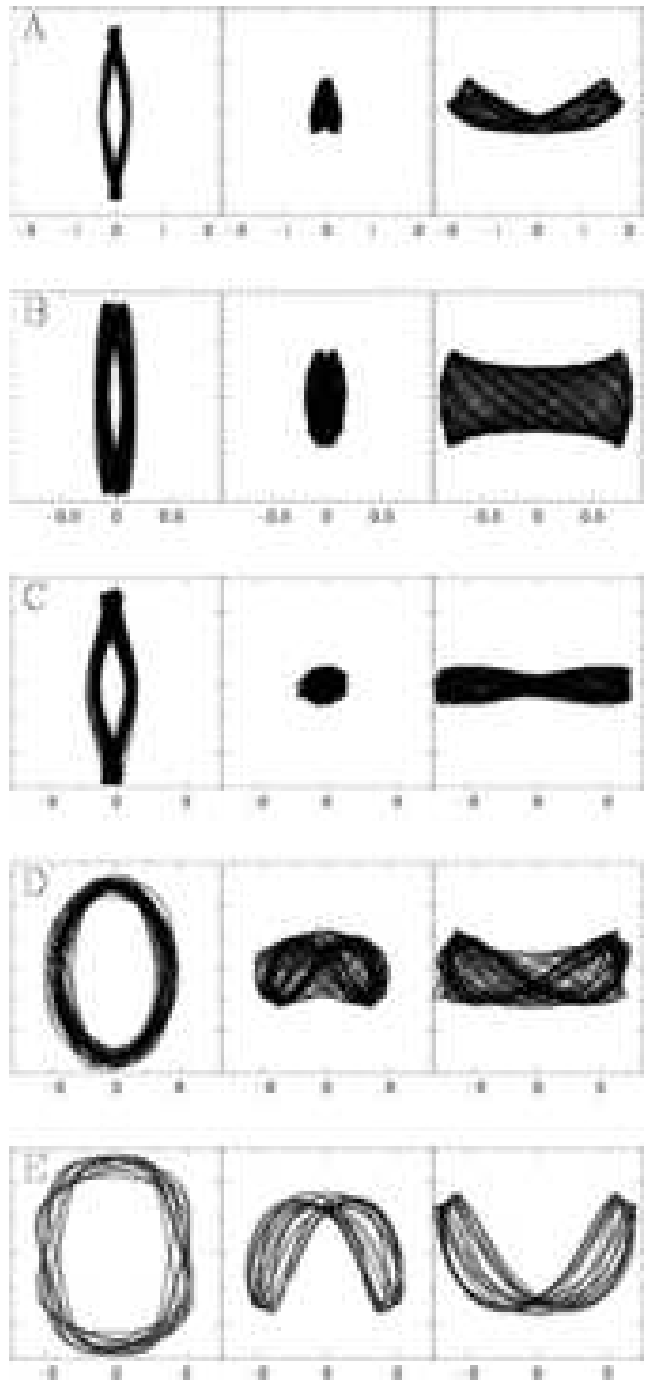}}
\end{center}
\caption[]{Six examples of typical halo near-ILR orbits. The ($x$, $y$),
  ($x$, $z$) and 
  ($y$, $z$) projections are shown in the left, middle and right
  panels, respectively. 
}
\label{fig:ILRorbits} 
\end{figure}

\begin{figure} 
\begin{center}
\rotatebox{0}{\includegraphics[width=18pc]{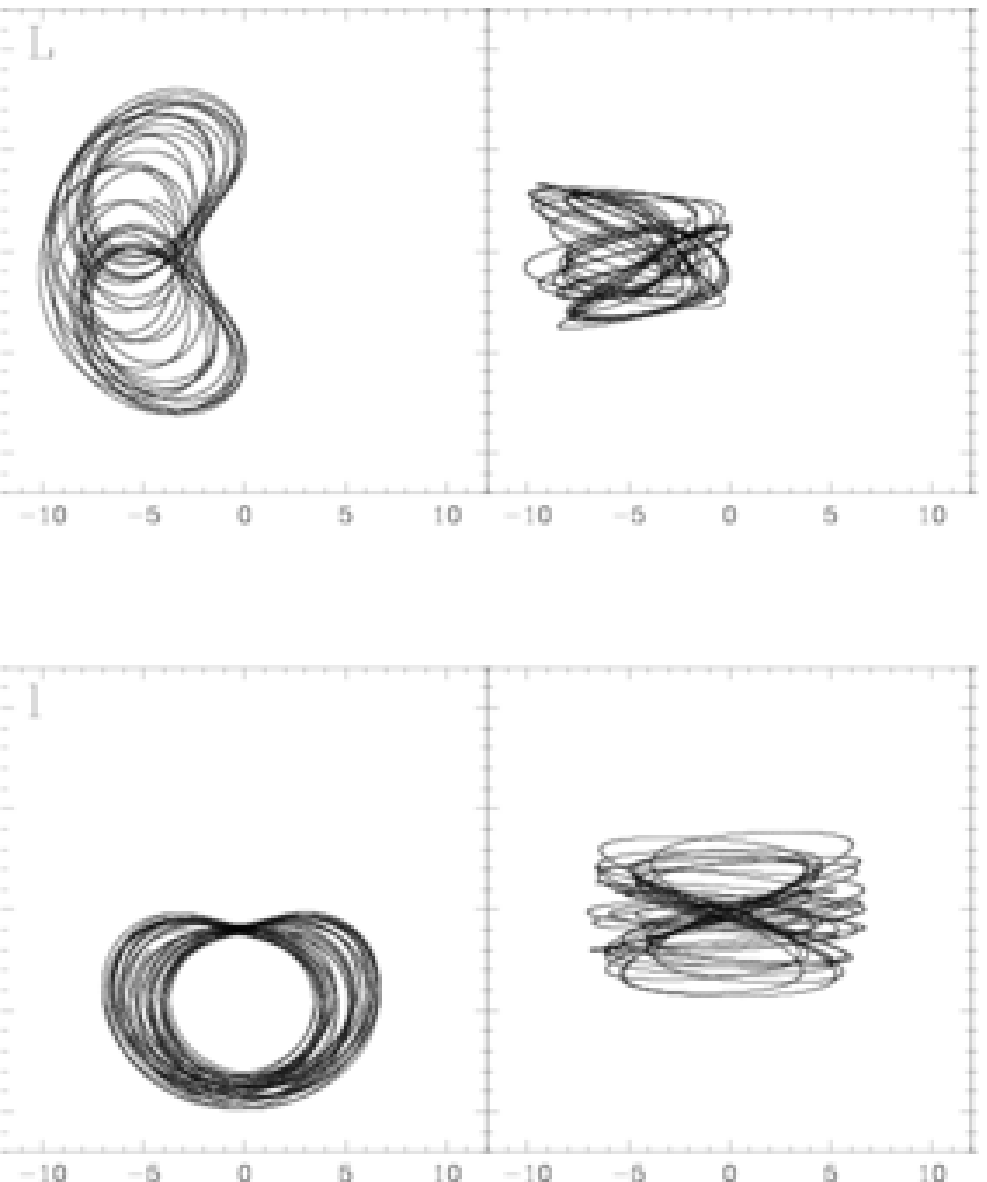}}
\end{center}
\caption[]{Two examples of typical halo near-CR orbits. The left
  panels give the face-on view, ($x$, $y$), and the right ones the
  end-on view, ($x$, $z$). 
}
\label{fig:CRorbits}
\end{figure}

Studies of orbital structure in 3D barred potentials can give useful
information on the families of periodic orbits, their stability and
the morphology of their orbits. They can not, however, give
information on how many orbits, if any, are trapped around a given
periodic orbit family. This information is only available from $N$-body
simulations. The orbital structure has been so far examined in
simulations of 2D or 3D discs (Sparke \& Sellwood 1987; Pfenniger \&
Friedli 1991; Berentzen et al. 1998), 
but never, so far, for their haloes. Yet, as discussed in sections
\ref{sec:theory} and \ref{sec:simul}, haloes also have
near-resonant orbits, and thus should have an interesting
orbital structure. I will here present some preliminary results on the
orbital structure of the disc and halo components. I use a method
which is, in several aspects, similar to that outlined in
the above mentioned papers. Namely, at a given time well after the bar
has grown, the potential and forces from a given simulation are
calculated on a grid, which is sometimes further made bisymmetric,
since most of 
the orbital families have that symmetry. The bar pattern speed at that
same time is also calculated from the simulation. In the above
mentioned papers, this information is used to calculate families of
periodic orbits. Unfortunately, numerical noise, which is inherent in
$N$-body simulations, limits such studies to few orbital families and
makes stability analyses very difficult, if not impossible. Instead of
this, I take the positions and velocities of 100\,000 disc and
100\,000 halo particles, 
drawn at random from the corresponding populations at the time at which the
potential and pattern speed are calculated. I use these as initial
conditions and follow the corresponding orbits for 40 bar
rotations. Examining these orbits leads to a number of conclusions,
some of which I will briefly discuss below. A more complete analysis,
comparing several simulations, will be given elsewhere (Athanassoula,
in preparation).

Of course, these orbits are not the same as those of the same particles
throughout the simulation. The simulation orbits start off in an
axially symmetric potential and evolve as the bar first forms and then 
evolves. During the formation phase the potential changes 
drastically with time, so that any results obtained in the way
described above should be considered with great caution. The formation
phase, however, is followed by a phase of calm secular evolution, in
which the potential changes slowly with time. It is during this phase
that the results obtained with the above method can be useful. The
evolution of the potential brings about a change in the orbital
structure. By studying this structure at a sequence of times, it is
possible to get insight on its evolution. It should, of course, always
be remembered that this is only an approximation, albeit the only one
known so far that can give useful results. This approximation is also
inherently present in any work on orbital structure, since such works
have so far used only non-evolving potentials, whether these are taken from
simulations, from observations, or from analytical models. 

\subsection{Morphology of halo near-resonant orbits}
\label{sec:examples}

Typical halo orbits trapped around the periodic orbits of
the x$_1$ tree are shown in figure~\ref{fig:ILRorbits}. They have
been chosen as representative of types of orbits frequently
encountered in the simulations presented here. The
plots are made in a frame of reference in which the bar is at rest and
the bar major axis is along the $y$ axis. 

Orbit A (upper
panels) must be trapped around a stable periodic orbit having 
two oscillations in the $z$ direction and two radial oscillations
for each rotation, i.e. an orbit of the x1v1 family,
presumably resembling that shown in the upper panels of 
figure~\ref{fig:3Dorbits}. Orbit B (second row) is
presumably trapped around an orbit of the x1v1 family {\it and} around its
symmetric with respect to the disc equatorial plane. This
orbit is located in the innermost part of the halo. 
 
Orbit C (third row) has a considerably smaller extent in the
$z$ direction. It is also more extended in the ($x$, $y$)
plane, so that it has a considerably smaller aspect ratio seen
side-on -- i.e. in the ($y$, $z$) projection -- than orbits A and B.
Its shape is also different. Orbits A and B
have a vertical extent which increases with
increasing distance from the center, as is the case for the periodic
orbits of the x1v1 family. Thus the highest point in $z$ is near the
maximum $y$ value, i.e. towards the tips of the ($x$, $y$) loop. On
the contrary, the orbit in the third row seen side-on appears
relatively flat for large $y$ values, i.e. for large $y$ values the
$z$-extent does not increase with distance from the center, but stays
roughly constant. 

Orbit C can have two possible origins. The first possibility is
that it is trapped around an x1v2 periodic orbit (see e.g. the last
row of panels of figure~\ref{fig:3Dorbits}). This family has members
with a similar orbital shape and extent to orbit C. It was, however, found to be
unstable in the models studied so far (Skokos et al. 2000a, b) and
thus cannot trap orbits around it. Changing the values of 
the parameters of the Skokos et al. models, it is possible to find cases which
have an x1v2 family with a stable section, but this is of a very short
extent in all the cases examined (Patsis, private
communication). Of course the potential in the simulations may not be
well described by one
of the models discussed by Skokos and collaborators and thus could
have considerable differences in its orbital stability properties. 

The second alternative is that orbits such as C are trapped around
periodic orbits of the x1v4 family. An example is given in the third
row of panels of figure~\ref{fig:3Dorbits}.
An orbit trapped around two periodic orbits of this family, which are
symmetric with respect to the equatorial plane, will look like orbit C
and have the right shape and extent. It will be possible to
distinguish between the two alternatives only after a complete study
of the orbital structure has been made. 

Orbits such as D, although rather frequent, are even more difficult to
classify. Seen face-on, such orbits show a simple oval, with no cusps or loops
at its tips. The edge-on views, however, reveal that the maximum $z$
displacement is not at the tips of the oval, but considerably
displaced. Again there are two possible alternatives as to the origin
of such orbits.

The first alternative is that this is due to a local twist of the
isodensities, which affects the edge-on, but not the face-on
view. Such a displacement is seen e.g. in the x1v4 periodic orbits
(third row of panels of figure~\ref{fig:3Dorbits}), even in cases where
the isodensities have no twist. 

The second alternative is
that orbit D is trapped around a member of a higher multiplicity family
bifurcating from the x$_1$, most probably one of multiplicity two. Such
periodic orbits close after two rotations and four radial oscillations
and thus can not be distinguished from those of the x$_1$ by their
frequency ratio. They had been found also in the models of Skokos et al., but
were not discussed at length since those papers concentrated on
periodic orbits of multiplicity 1. Orbit E, in the lowest row of panels
of figure~\ref{fig:ILRorbits}, argues yet stronger about its link with a
periodic orbit of multiplicity 2. Such orbits are rather frequent
amongst the ones having a 1:2 frequency ratio and this argues that families with
multiplicity higher than 1 should be considered in future orbital
studies, 

The orbits in figure~\ref{fig:CRorbits} are `banana-like'. The one
in the first row orbits around the $L_4$, or $L_5$, Lagrangian point
and must be trapped around a 
long period banana orbit (Contopoulos \& Grosb{\o}l 1989). The example in
the second row orbits around the $L_1$, or $L_2$, Lagrangian
point. Although these points are known to be unstable (e.g. Binney \&
Tremaine 1987), Skokos et al. found a family whose members orbit around
$L_1$, or $L_2$, and has considerable stable parts, They called it $l_1$. The
morphology of the orbits of this 
family resembles that of the short period orbits around $L_4$, or
$L_5$, rotated by $\pi$/2. The shape and orientation of orbit $l$
(second row of figure~\ref{fig:CRorbits}) shows that it must be trapped
around a periodic orbit of the $l_1$ family.  

In general, the shapes of the orbits of the halo near-resonant
particles are similar to the corresponding disc ones. Although
this might seem strange at first glance, it should in fact have been
expected, since the periodic orbits are characteristic of the total
potential, i.e. they are the same for the disc and the halo. {
   
\subsection{Fraction of chaotic orbits}
\label{sec:chaos}

\begin{figure}
\begin{center}
\includegraphics{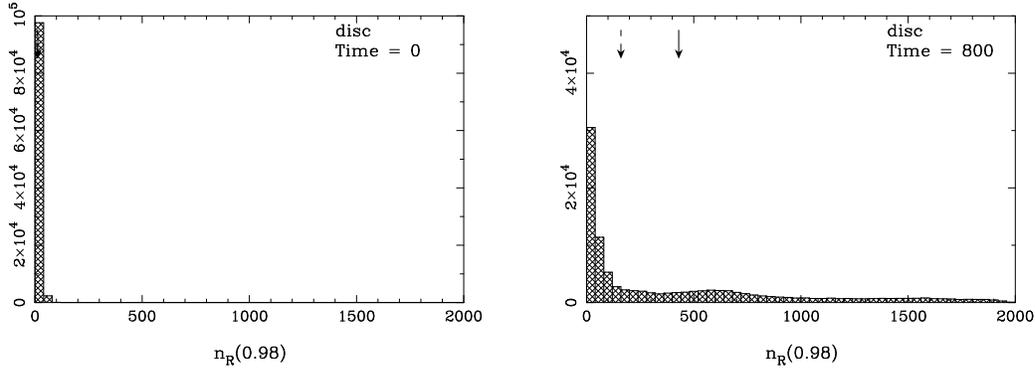} 
\end{center}
\vspace{6.5cm}
\caption{Number of disc orbits as a function of the complexity
  index $n_R(0.98)$. The left panel corresponds to the initial time
  and shows that all orbits are regular, i.e. have small 
  complexity. The right panel corresponds to a time  after a strong
  bar has grown and shows that many orbits have acquired very high
  complexity, i.e. have become chaotic. The solid-line and
  dashed-line-arrows give the mean and the median values,
  respectively.
}
\label{fig:chaos-histo}
\end{figure}

\begin{figure}
\begin{center}
\includegraphics{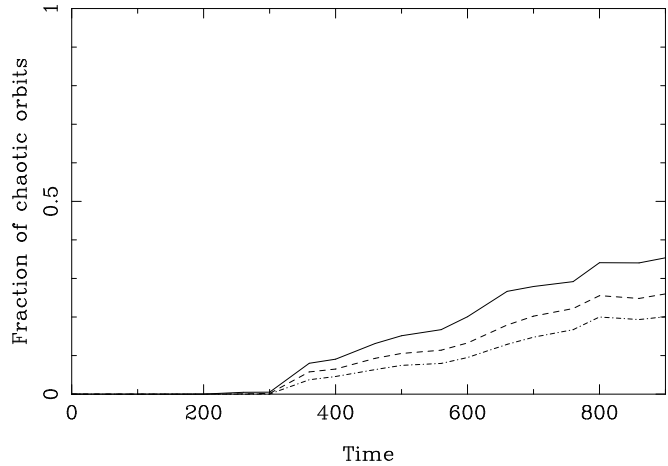} 
\end{center}
\vspace{7.7cm}
\caption{Fraction of the disc orbits that have a complexity above
  a given threshold as a function of time for a simulation with a
  strong bar. The three thresholds are 100 (solid line), 150
  (dashed line) and 200 (dot-dashed line).
}
\label{fig:chaos-time}
\end{figure}

The amount of chaos as a function of bar strength has
already been addressed by Athanassoula et al. (1983), with the help of
surfaces of sections. These authors plotted the fraction of the
area of the surfaces of section which is covered by chaotic orbits 
as a function of the bar mass and axial ratio. They found that there
is a clear trend, in the sense that stronger bars have a larger part of their
surfaces of section covered by chaotic orbits. This was confirmed
by Teuben \& Sanders (1985), for a different model. Although
very indicative, these results do not give a clear estimate of the
amount of chaos in barred galaxies. Indeed, they do not take into
account the time spent by an orbit between two consecutive crossings
of the surface of section (e.g. Binney, Gerhard \& Hut 1985) and, most
important, they have 
no way of determining whether a given regular or chaotic orbit is
populated, or not. 

This information can only be obtained from $N$-body simulations. 
I have thus calculated the complexity of 100\,000 disc particles, and
100\,000 halo particles, taken at random from the corresponding
population, and that for several times 
during the evolution. Initially, by construction, both the disc and the
halo are axisymmetric, and there is no chaos in either component. This
is, however, introduced gradually as the bar first forms and then
becomes stronger with time, so that towards the end of the simulation
there is a considerable fraction of chaotic orbits. This is seen, for
the disc orbits, in figure~\ref{fig:chaos-histo}. This shows, in form of
a histogram, the number of disc particles as a function of their
complexity, both initially and for a time towards the end of the
simulations. We note that the distribution acquires a considerable
tail towards complex/chaotic orbits.

The evolution of the number of disc chaotic orbits with time is further
examined in figure~\ref{fig:chaos-time}. 
For this, I plot the number of orbits with complexity above a
given threshold
as a function of time. Before the bar forms there are no chaotic orbits, as
expected. As the bar grows, the amount of chaos increases and,
at the end of the simulation, nearly one third of the
disc orbits have a complexity higher than 100. 

\indent
\subsection{Where were the halo near-resonant particles initially?}
\label{sec:resstart}

\indent
Where do the halo particles at (near-) resonance
originate from? Do all particles have equal probability of getting trapped at
a resonance during the evolution, or are there preferred targets? To
answer this question I first found the
particles which are at (near-) resonance at a time well after the bar has
formed and then traced them back to the initial conditions. I could
thus compare the distribution of their
main properties (cylindrical and spherical radius, distance from the
equatorial plane, $z$ component of the velocity and of the angular
momentum, kinetic energy etc)  with that of all the particles in the
halo, both taken at $t$ = 0. For CR, which is well populated and for 
which it is thus easy to make statistical tests, a
Kolmogorov-Smirnov test shows beyond doubt that the two populations
are not identical, i.e. that the particles which at later times are at
CR are not initially randomly chosen from the initial halo population. The
particles that will later get trapped at CR do not come initially from
the innermost and outermost regions. Instead,
they have preferentially intermediate cylindrical and
spherical radii. They have preferentially
(absolutely) smaller values of $v_z$ and considerably larger values of $J_z$.

\section{The effect of a central mass concentration on a bar}

\begin{figure} 
\begin{center}
\rotatebox{0}{\includegraphics[width=25pc]{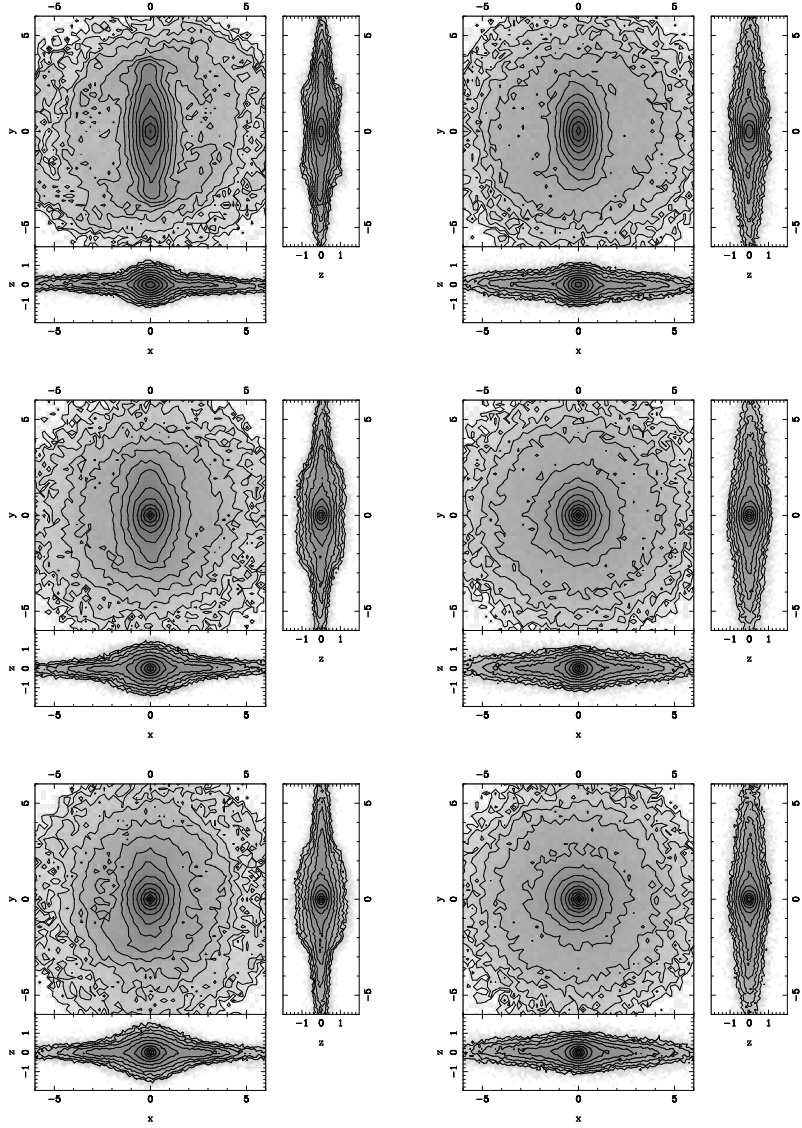}}
\end{center}
\caption[]{Effect of a CMC. The left panels correspond to MH-type models
  and the right ones to MD-type models. The upper panel shows the
  disc component at the time the CMC is introduced and the other two at
  $\Delta T = $ 300 later. The mass of the
  CMC is 0.05 for the middle panel and 0.1 for the lower one. Each
  sub-panel shows one of the three orthogonal views of the disc
  component. The right sub-panel gives the edge-on 
  side-on view, the lower left one gives the edge-on end-on view and
  the upper left one gives the face-on view. The projected density of the
  disc is given by 
  grey-scale and also by isocontours (spaced logarithmically).}
\label{fig:bhg_basic}
\end{figure}

Central mass concentrations (hereafter CMCs) can be hazardous for the
growth and even for the existence of a bar. This was first discussed by
Hasan \& Norman (1990) and Hasan, Pfenniger \& Norman (1993),
who studied the orbital structure in a 
rigid potential with both a bar and a CMC. They showed
that the CMC alters the stability of the x$_1$ orbits, making them
largely unstable and thus came to the conclusion that
a CMC can destroy the bar if it is sufficiently massive and/or
sufficiently centrally concentrated. This work was extended with
the help of $N$-body simulations first by
Norman, Sellwood \& Hasan (1996) and later by Hozumi \& Hernquist
(1998, 1999, 2004) and by Shen \& Sellwood (2004). All three groups
used $N$ body simulations with live discs and rigid haloes. In all
cases the CMC was introduced gradually, in order to avoid transients. 

\begin{figure*}
\setlength{\unitlength}{2cm}
\begin{center}
\includegraphics[scale=0.45,angle=-90.]{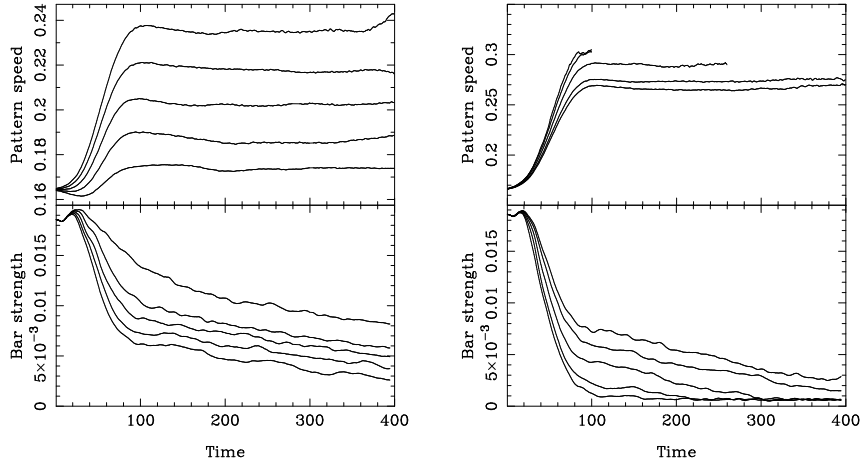}
\end{center}
\caption{Pattern speed of the bar as a function of time (upper
  panels). For comparison, I plot in the lower panels a
  measure of the bar strength (the maximum value of the amplitude of
  the $m$ = 2 component of the mass distribution), also as a function
  of time. The left panels refer to a series of 
  simulations with different CMC mass (from top to bottom in the lower
  panel and from bottom to top in the upper one : 0.01, 0.02,
  0.03, 0.04, 0.05; all with a CMC radius of 0.01), while the right
  panels correspond to a series   
  of simulations with different CMC radius (from top to bottom in the upper
  panel and from bottom to top in the lower one : 0.01, 0.02, 0.05, 0.08, 0.1;
  all with a CMC mass of 0.1). In all cases the CMC is introduced
  gradually in the first 100 time units. 
} 
\label{fig:omegap_Sb}
\end{figure*}
 
Seen the important role that the response of the halo can play on the
evolution of barred galaxies, Athanassoula, Lambert \& Dehnen (2005) 
revisited this problem, now using a live halo, and I will here
summarise some of their results. They find that the effect of the CMC
depends drastically on  
the model. This is illustrated in figure~\ref{fig:bhg_basic}, which
compares the effect of identical CMCs (of mass 0.05 and radius 0.01 for
the middle panels, and for mass 0.1 and radius 0.01 for the lower panels)
on two different barred galaxy models. The left column corresponds to
an MH-type  
model, and the right one to an MD-type model. The difference is quite
important. These CMCs decrease the strength of the bar in the MH-type
models, but fully destroy it in MD-type ones. This figure also shows
that the lowering of the bar strength is due to a decrease of the bar
length and to more axisymmetric innermost parts. The latter can be
understood since this is the vicinity of the CMC.  

This difference between MH-type and MD-type models could be due to the
role of the halo in the two cases. Indeed, in the MH-type haloes the
inner resonances in the halo are more populated, so that the halo can absorb
more angular momentum, compared to MD-type haloes (A03). This extra
angular momentum is taken from the bar and will, as discussed in
sections~\ref{sec:theory} and \ref{sec:simul}, tend to increase its strength and
will work against the CMC, whose effect will thus be lessened. This
is indeed what the simulations of Athanassoula et al. (2005) show.

The CMC also affects the vertical structure of the bar. 
Figure~\ref{fig:bhg_basic} shows that in MH-type models the peanut
initially seen in
the side-on view is converted to a boxy or elliptical-like
shape. The radial extent, however, of this vertically protuberant
part stays roughly the same. The CMC in the MD-type bar destroys the
boxiness of the side-on view.

The CMC also causes an increase of the bar pattern speed. This is not
always easy to measure, since, due to the CMC, the bar amplitude is
severely decreased.  
It is, nevertheless, clear for the two sequences of simulations, compared
in figure~\ref{fig:omegap_Sb}. This effect is in agreement with the
anti-correlation between the bar strength and pattern speed, found
in A03.

\vskip .2in

\noindent
{\bf Acknowledgments}
It is a pleasure to thank many friends and colleagues for interesting
and motivating discussions, in particular 
A. Bosma, M. Bureau, G. Contopoulos, W. Dehnen, A. Misiriotis,
P. Patsis, M. Tagger, N. Voglis, 
G. Aronica, I. Berentzen, K. Freeman, B. Fuchs, K. Holley-Bockelmann,
A. Kalnajs, A. Klypin, J. Kormendy, D. Lynden-Bell, F. Masset, I. Shlosman
Ch. Skokos and M. Weinberg. I thank Jean-Charles
Lambert for his invaluable help with the simulation software and the
administration of the runs and W. Dehnen for 
making available to me his tree code and related programs. 
I also thank the Observatoire de Marseille, the region PACA, the
INSU/CNRS and the University of Aix-Marseille I for funds to develop
the computing facilities used for the calculations in this paper.
\vskip .2in

\noindent
{\bf References}
\def\rr{\par\noindent\parshape=2 0cm 8cm 1cm 7cm}

\noindent
{Athanassoula, E., {\em Monthly Notices of the Royal Astronomical Society},
 259, 365, 1992.}

\noindent\hangindent=20pt
Athanassoula, E. 
\newblock In R. Buta, D.A. Crocker and B.G. Elmegreen 
{\em Barred Galaxies}, pages 309,
  Astron. Soc. Pac. Conf. Series 91, 1996.

\noindent\hangindent=20pt
Athanassoula, E.
\newblock {\em Astrophysical Journal Letters}, 569, L83, 2002 (A02).

\noindent\hangindent=20pt
Athanassoula, E.
\newblock {\em Monthly Notices of the Royal Astronomical Society},
341, 1179, 2003a (A03). 

\noindent\hangindent=20pt
Athanassoula, E. 
\newblock In J. Makino and P. Hut, editors, {\em
Astrophysical supercomputing using particles},
Astron. Soc. Pac. Conference Series, IAU Symp., 208, 177, 2003b.

\noindent\hangindent=20pt
Athanassoula, E.
\newblock {\em Celestial Mechanics and Dynamical Astronomy}, 45,
9, 2005. 

\noindent\hangindent=20pt
Athanassoula, E., Bienaym\'e, O., Martinet, L., Pfenniger, D. 
\newblock {\em Astronomy and Astrophysics} 127, 349, 1983.

\noindent\hangindent=20pt
Athanassoula, E., Bosma, A.,  Mujica, R.  
\newblock {\em Disks of
  Galaxies : Kinematics, Dynamics and Perturbations}, 
  Astron. Soc. Pac. Conf. Series 275, 2002.

\noindent\hangindent=20pt
Athanassoula, E., Lambert, J. C., Dehnen, W.
\newblock {\em Monthly Notices of the Royal Astronomical Society}, 
submitted, 2005.

\noindent\hangindent=20pt
Athanassoula, E., Misiriotis, A. 
\newblock {\em Monthly Notices of the Royal Astronomical Society}, 
330, 35, 2002 (AM02).

\noindent\hangindent=20pt
Athanassoula, E., Morin, S., Wozniak, H., Puy, D., Pierce, M. J.,
Lombard, J., Bosma, A.
\newblock {\em Monthly Notices of the Royal Astronomical Society}, 
245, 130, 1990.

\noindent\hangindent=20pt
Athanassoula, E., Sellwood, J. A. 
\newblock {\em Monthly Notices of the Royal Astronomical Society},
221, 213, 1986. 

\noindent\hangindent=20pt
Berentzen, I., Athanassoula, E., Heller, C. H. ,Fricke, K. J. 
\newblock {\em Monthly Notices of the Royal Astronomical Society}, 
347, 220, 2004.

\noindent\hangindent=20pt
Berentzen, I., Heller, C. H. Shlosman I. \& Fricke, K. J. 
\newblock {\em Monthly Notices of the Royal Astronomical Society}, 
300, 49, 1998.

\noindent\hangindent=20pt
Binney, J. 
\newblock {\em Monthly Notices of the Royal Astronomical Society},
201, 1, 1982.

\noindent\hangindent=20pt
Binney, J., Gerhard, O., Hut, P. 
\newblock {\em Monthly Notices of the Royal Astronomical Society},
215, 59, 1985.

\noindent\hangindent=20pt
Binney, J., Tremaine, S. 
\newblock {\em Galactic Dynamics}, Princeton Univ. Press, 1987.

\noindent\hangindent=20pt
Block, D., Freeman, K. C., Puerari, I., Groess, R., Block, L.  
\newblock {\em Penetrating bars through masks of cosmic dust: The Hubble
  Tuning Fork Strikes a New Note}, 
  Kluwer publ., in press, 2005.

\noindent\hangindent=20pt
Bosma, A. 
\newblock In G. Longo, M. Capaccioli and G. Busarello 
{\em Morphological and Physical Classification of Galaxies}, p. 207,
  Kluwer pub., 1992.

\noindent\hangindent=20pt
Bottema, R. 
\newblock {\em Monthly Notices of the Royal Astronomical Society},
344, 358, 2003.

\noindent\hangindent=20pt
Buta, R.J.
\newblock {\em Astrophysical Journal Supplements}, 61, 609, 1986.

\noindent\hangindent=20pt
Buta, R., Crocker, D.A., Elmegreen, B.G.
\newblock {\em Barred Galaxies}, Astron. Soc. Pac. Conf. Series 91, 1996.

\noindent\hangindent=20pt
Chandrasekhar, S. 
\newblock {\em Astrophysical Journal}, 97, 255, 1943.

\noindent\hangindent=20pt
Contopoulos, G.
\newblock {\em Astronomy and Astrophysics}, 81, 198, 1980.

\noindent\hangindent=20pt
Contopoulos, G.
\newblock {\em Order and chaos in Dynamical Astronomy}, Springer, 2002.

\noindent\hangindent=20pt
Contopoulos, G., Grosb{\o}l, P. 
\newblock {\em Astronomy and Astrophysics Reviews}, 1, 261, 1989.

\noindent\hangindent=20pt
Contopoulos, G., Papayannopoulos, T.
\newblock {\em Astronomy and Astrophysics}, 92, 33, 1980.

\noindent\hangindent=20pt
Debattista, V. P., Sellwood, J. A. 
\newblock {\em Astrophysical Journal}, 543, 704, 2000.

\noindent\hangindent=20pt
Efstathiou, G., Lake, G., Negroponte, J. 
\newblock {\em Monthly Notices of the Royal Astronomical Society},
199, 1069, 1982.

\noindent\hangindent=20pt
Elmegreen, B.G., Elmegreen, D.M. 
\newblock {\em Astrophysical Journal}, 288, 438, 1985.

\noindent\hangindent=20pt
Elmegreen, D.M., Elmegreen, B.G., Bellin, A.D. 
\newblock {\em Astrophysical Journal}, 364, 415, 1990.

\noindent\hangindent=20pt
El-Zant, A., Shlosman, I. 
\newblock {\em Astrophysical Journal}, 577, 626, 2002.

\noindent\hangindent=20pt
El-Zant, A., Shlosman, I. 
\newblock {\em Astrophysical Journal}, 595, L41, 2003.

\noindent\hangindent=20pt
Eskridge et al.
\newblock {\em Astronomical Journal}, 119, 536, 2000.

\noindent\hangindent=20pt
Ferrers, N. M.
\newblock {\em Q. J. Pure Appl. Math.}, 14, 1, 1877.

\noindent\hangindent=20pt
Gadotti, D. A., de Souza, R. E. 
\newblock {\em Astrophysical Journal}, 583, L75, 2003.

\noindent\hangindent=20pt
Grosb{\o}l P., Pompei, E., Patsis, P. A. 
\newblock In E. Athanassoula, A. Bosma and R. Mujica {\em Disks of
  galaxies : Kinematics, dynamics and perturbations}, 305,
  Astron. Soc. Pac. Conf. Series 275, 2002.

\noindent\hangindent=20pt
Hasan, H., Norman, C. 
\newblock {\em Astrophysical Journal}, 361, 69, 1990. 

\noindent\hangindent=20pt
Hasan, H., Pfenniger, D., Norman, C. 
\newblock {\em Astrophysical Journal}, 409, 91, 1993. 

\noindent\hangindent=20pt
Hernquist, L., Weinberg, M.  
\newblock {\em Astrophysical Journal}, 400, 80, 1992. 

\noindent\hangindent=20pt
Hohl, F. 
\newblock {\em Astrophysical Journal}, 168, 343, 1971. 

\noindent\hangindent=20pt
Holley-Bockelmann, K., Weinberg, M. D., Katz, N.  
\newblock {\em astro-ph/0306374}, 2003.

\noindent\hangindent=20pt
Hozumi, S., Hernquist, L.  
\newblock {\em astro-ph 9806002},  1998. 

\noindent\hangindent=20pt
Hozumi, S., Hernquist, L.  
\newblock In D. Merritt, J. A. Sellwood and
M. Valluri, {\em Galaxy Dynamics}, Astron. Soc. Pac. Conf. Series 182, 259, 1999. 

\noindent\hangindent=20pt
Hozumi, S., Hernquist, L.  
\newblock {\em Publications of the Astronomical Society of Japan},
submitted 2004.  

\noindent\hangindent=20pt
Kalnajs, A. J.  
\newblock {\em Astrophysical Journal}, 166, 275, 1971. 

\noindent\hangindent=20pt
Kandrup, H. E., Eckstein, B. L., Bradley, B. O. 
\newblock {\em Astronomy and Astrophysics}, 320, 65, 1997.

\noindent\hangindent=20pt
Kormendy, J. 
\newblock In L. Martinet and M. Mayor {\em Morphology and Dynamics of Galaxies},
  Geneva Obs. publ., Geneva, 1982.

\noindent\hangindent=20pt
Little, B., Carlberg, R. G. 
\newblock {\em Monthly Notices of the Royal Astronomical Society}, 250, 161, 1991a.

\noindent\hangindent=20pt
Little, B., Carlberg, R. G. 
\newblock {\em Monthly Notices of the Royal Astronomical Society}, 251, 227, 1991b.

\noindent\hangindent=20pt
Lynden-Bell, D., Kalnajs, A. J. 
\newblock {\em Monthly Notices of the Royal Astronomical Society}, 
157, 1, 1972.

\noindent\hangindent=20pt
Michalodimitrakis, M. 
\newblock {\em Astrophysics and Space Science}, 33, 421, 1975. 

\noindent\hangindent=20pt
Miller, R. H., Prendergast, K. H., Quirk, W. J. 
\newblock {\em Astrophysical Journal}, 161, 903, 1970. 

\noindent\hangindent=20pt
Norman, C., Sellwood, J. A., Hasan, H.
\newblock {\em Astrophysical Journal}, 462, 114, 1996. 

\noindent\hangindent=20pt
Ohta, K. 
\newblock In R. Buta, D. A. Crocker and B. G. Elmegreen 
{\em Barred Galaxies}, Astron. Soc. Pac. Conf. Series 91, 37, 1996.

\noindent\hangindent=20pt
Ohta, K., Hamabe, M., Wakamatsu, K. 
\newblock In R. Buta, D.A. Crocker and B.G. Elmegreen 
{\em Barred Galaxies}, Astron. Soc. Pac. Conf. Series 91, 37, 1990.

\noindent\hangindent=20pt
O'Neill, J. K., Dubinski, J. 
\newblock {\em Monthly Notices of the Royal Astronomical Society},
346, 251, 2003. 

\noindent\hangindent=20pt
Ostriker, J. P., Peebles, P. J. E. 
\newblock {\em Astrophysical Journal}, 186, 467, 1973. 

\noindent\hangindent=20pt
Patsis, P. A., Athanassoula, E., Quillen, A.
\newblock {\em Monthly Notices of the Royal Astronomical Society},
 483, 731, 1997. 

\noindent\hangindent=20pt
Patsis, P. A., Skokos, Ch., Athanassoula, E.
\newblock {\em Monthly Notices of the Royal Astronomical Society},
 337, 578, 2002. 

\noindent\hangindent=20pt
Patsis, P. A., Skokos, Ch., Athanassoula, E.
\newblock {\em Monthly Notices of the Royal Astronomical Society}, 
342, 69, 2003.

\noindent\hangindent=20pt
Pfenniger, D.
\newblock {\em Astronomy and Astrophysics}, 134, 373, 1984a.

\noindent\hangindent=20pt
Pfenniger, D.
\newblock {\em Astronomy and Astrophysics}, 141, 171, 1984b.

\noindent\hangindent=20pt
Pfenniger D. Friedli D. 
\newblock {\em Astronomy and Astrophysics}, 252, 75, 1991.

\noindent\hangindent=20pt
Sandage, A.
\newblock {\em The Hubble Atlas of Galaxies}, Carnegie Inst. Publ.,
Washington, 1961. 

\noindent\hangindent=20pt
Shen, J., Sellwood, J. A.
\newblock {\em Astrophysical Journal}, 604, 614, 2004.

\noindent\hangindent=20pt
Skokos, Ch., Patsis, P. A., Athanassoula, E. 
\newblock {\em Monthly Notices of the Royal Astronomical Society}, 
333, 847, 2002a.

\noindent\hangindent=20pt
Skokos, Ch., Patsis, P. A., Athanassoula, E. 
\newblock {\em Monthly Notices of the Royal Astronomical Society}, 
333, 861, 2002b.

\noindent\hangindent=20pt
Sparke, L. S. Sellwood, J. A. 
\newblock {\em Monthly Notices of the Royal Astronomical Society},
225, 653, 1991. 

\noindent\hangindent=20pt
Teuben, P. J., Sanders, R. H.
\newblock {\em Monthly Notices of the Royal Astronomical Society}, 
212, 257, 1985.

\noindent\hangindent=20pt
Toomre, A.
\newblock {\em Annual Reviews of Astronomy \& Astrophysics}, 15, 437, 1977.

\noindent\hangindent=20pt
Tremaine, S., Weinberg, M. D.
\newblock {\em Monthly Notices of the Royal Astronomical Society}, 
209, 729, 1984.

\noindent\hangindent=20pt
Valenzuela, O., Klypin, A. 
\newblock {\em Monthly Notices of the Royal Astronomical Society},
345, 406, 2003. 

\noindent\hangindent=20pt
de Vaucouleurs, G., Freeman, K. C.
\newblock {\em Vistas in Astronomy}, 1, 163, 1972.

\noindent\hangindent=20pt
Weinberg, M. D.
\newblock {\em Monthly Notices of the Royal Astronomical Society}, 
213, 451, 1985.

\noindent\hangindent=20pt
Weinberg, M. D.
\newblock {\em Monthly Notices of the Royal Astronomical Society},
astro-ph/0404169, 2004.

\end{document}